\documentclass[]{JHEP3}   

\JHEPspecialurl{http://jhep.sissa.it/JOURNAL/JHEP3.tar.gz}

\usepackage{epsfig,multicol}

\newcommand\fverb{\setbox\pippobox=\hbox\bgroup\verb}
\newcommand\fverbdo{\egroup\medskip\noindent%
            \fbox{\unhbox\pippobox}\ }
\newcommand\fverbit{\egroup\item[\fbox{\unhbox\pippobox}]}
\newbox\pippobox

\newcommand{\lsim}{\mathrel{\raisebox{-.6ex}{$\stackrel{\textstyle<}{\sim}$}}}
\newcommand{\gsim}{\mathrel{\raisebox{-.6ex}{$\stackrel{\textstyle>}{\sim}$}}}
\newcommand{\be}{\begin{eqnarray}}
\newcommand{\ee}{\end{eqnarray}}

\title{
\begin{flushright}
\normalsize{ FERMILAB-PUB-07-211-T\\FTUV 07-0618}
\end{flushright}
Quintessence, inflation and baryogenesis\\
from a single pseudo-Nambu-Goldstone boson}

\author{       {Gabriela Barenboim}\\
\normalsize\emph{Departament de F\'isica Te\`orica and IFIC, Universitat de
Val\`encia - CSIC}\\
\emph{Carrer Dr. Moliner 50, E-46100 Burjassot (Val\`encia), Spain}\\
Email: \email{gabriela.barenboim@uv.es}\\}

\author{\textbf{Joseph D. Lykken}\\
\normalsize\emph{Fermi National Accelerator Laboratory}\\
\emph{P.O. Box 500, Batavia, IL 60510, USA}\\
Email: \email{lykken@fnal.gov}}

\abstract{
We exhibit a model in which
a single pseudo-Nambu-Goldstone boson explains dark energy, 
inflation and baryogenesis.
The model predicts correlated signals in future collider experiments,
WIMP searches, proton decay experiments, dark energy probes, and the
PLANCK satellite CMB measurements.
}


\begin{document}


\section{Introduction}
The most plausible candidate for a quintessence explanation
of dark energy is a pseudo-Nambu-Goldstone boson (PNGB)
of a spontaneously
and explicitly broken global $U(1)$ 
symmetry \cite{Frieman:1991tu}\cite{Frieman:1995pm}.
Denoting the scale of the explicit breaking by $M$ and
the scale of the spontaneous breaking by $f$, the
effective mass of the PNGB is $m_{\rm eff} \sim M^2/f$. The
required value of this mass for a successful quintessence
model is of order the Hubble parameter $H_0 \sim 10^{-33}$ eV;
this can be obtained naturally if the spontaneous breaking
scale $f$ is very large, roughly comparable to the Planck scale
$M_{\rm Planck} \simeq 10^{19}$ GeV. Because of the symmetry,
the PNGB has only derivative couplings to matter and radiation,
plus couplings whose dimensionless strength is suppressed
by powers of $M/f \sim 10^{-22}$. This evades a number of strong experimental
and observational constraints on weakly-coupled ultralight
scalars \cite{Carroll:1998zi}. 

In a more general class of PNGB quintessence models,
the effective mass of the PNGB will vary over time,
\textit{i.e.} it will be a function of the scale factor
$a$ obtained by solving the coupled cosmological equations
of motion. Since $f$
is comparable to the Planck scale, it is possible that
$m_{\rm eff}$ was much larger at early times, without losing
the key property that $m_{\rm eff}/f \ll 1$.
This raises the possibility that the quintessence PNGB
field may also have been responsible for primordial inflation,
albeit in some modification of the usual scenario.

A simple avenue towards quintessential inflation is
then to assume that $m_{\rm eff}$ and the PNGB scalar potential
$V$ scale (at least roughly) like some power of the Hubble rate $H$.
Any model with $m_{\rm eff} \sim H$ has the additional virtue of removing the
coincidence problem, \textit{i.e.} the fact that the ratio of
scales $M/f$ is of order $H_0$ is no longer a coincidence, but rather
has some dynamical origin.

Any model of quintessential inflation must explain why
the energy density of the universe was dominated by
radiation at the time of Big Bang Nucleosynthesis (BBN),
even though the inflaton dominates the energy density now
and dominated it in a primordial epoch as well. For a
PNGB, the simplest explanation is that the PNGB decays
to matter and radiation via derivative couplings of the
form
\be\label{eqn:firstmattercoup}
\lambda_{ij}\frac{f}{M_{\rm Planck}}\,\;g^{\mu\nu}\partial_{\mu}  
\theta \,\bar{\psi}_i \gamma_{\nu} \psi_j \; ,
\ee 
where $\theta$ denotes the PNGB field rescaled
by $f$ to make it dimensionless, and the $\lambda_{ij}$
are moderately small dimensionless couplings.
For $f \sim M_{\rm Planck}$, this will allow large
entropy production from PNGB decays. Since the PNGB
potential varies over time, the equation of state of the
PNGB also varies. Thus it is natural to have periods
of inflation interspersed with periods of radiation
dominance.

The matter coupling (\ref{eqn:firstmattercoup}) was
introduced by Cohen and Kaplan in their thermodynamic
model for baryogenesis \cite{Cohen:1987vi}.  
In their scenario $B$ or $L$
violating processes occur via dimension six four-fermion
operators suppressed by a relatively low 
scale $\Lambda \sim 10^{8}$ GeV. Combined with a PNGB
possessing dimension five couplings like (\ref{eqn:firstmattercoup}),
they generate a baryon asymmetry in thermal equilibrium
(at temperatures $\gsim 10^{8}$ GeV), and a further
asymmetry at lower temperatures from PNGB decays.

In this paper we
construct a natural model of PNGB quintessential inflation
that also implements the Cohen-Kaplan mechanism for
baryogenesis. At the same time that we provide quintessence
and inflation, our model gives a simpler explanation
of the baryon asymmetry than the original
scenario of \cite{Cohen:1987vi}. We are able to
assume that the scale for the dimension six $B$, $L$
violating operators is of order $10^{15}$ GeV,
or of order $10^{11}$ GeV for purely $L$ violating
operators. Thus we are slightly above or saturating
the current experimental bounds from proton decay \cite{Nath:2006ut}. 
Our cosmological evolution begins with generic
initial conditions, unlike the original models
of \cite{Frieman:1991tu}\cite{Frieman:1995pm},
at an initial temperature $T_0 \sim 10^{17}$ GeV,
which is comfortably less than $f \sim M_{\rm Planck}$.
Our model requires no unnatural tunings other than
that of the cosmological constant, a tuning that is unavoidable
since quintessence does not solve the cosmological
constant problem.

\section{FRW cosmology driven by a Nambu-Goldstone boson}

Consider a theory with a $U(1)$ global symmetry under which some
complex scalar field $\Phi (x,t)$ transforms as
\be
\Phi \to {\rm e}^{i\alpha}\,\Phi \; ,
\ee
where $\alpha$ is a constant.
Other fields, including fermions, may also transform nontrivially;
in particular the symmetry may be a chiral symmetry.
We imagine that this symmetry is spontaneously broken at some
high scale, near $M_{\rm Planck}$, determined by the vev $f$ of the scalar.
Expanding around this vev gives
\be
\Phi = (f + \sigma(x,t))\,{\rm e}^{i\theta(x,t)}  \; ,
\ee
where $\sigma$ is a real scalar and $\theta$ is another real scalar which
has been rescaled by $1/f$ to be dimensionless. The effective theory at lower
energies has the original symmetry nonlinearly realized, with the
Nambu-Goldstone boson (NGB) $\theta$ undergoing a shift:
\be\label{eqn:shift}
\theta(x,t) \to \theta(x,t) + \alpha  \; .
\ee
Obviously the NGB only has derivative couplings in this effective theory.
Higher dimension operators are suppressed by powers of a high scale $f$
which we are roughly equating to $M_{\rm Planck}$. To leading order in
$f$ the NGB action is just
\be
\int d^4x \, \sqrt{-g}\; \frac{1}{2}f^2\, g^{\mu\nu}\partial_{\mu}\theta
\partial_{\nu} \theta
\; .
\ee
Let us further suppose that the energy density of the universe
during some epoch is dominated by the NGB. Then the Friedmann
equation is approximately
\be\label{eqn:NGB1}
H^2 = {f^2\over 3k^2} \dot{\theta}^2 \;,
\ee
where $k^2 = M_{\rm Planck}^2/4\pi$. The equation of motion for $\theta$
(again to leading order in $f$) is just
\be\label{eqn:NGB2}
\ddot{\theta} + 3H\dot{\theta} = 0 \;,
\ee
while the continuity equation is given by
\be\label{eqn:NGB3}
\dot{\rho}_{\theta} = -6H\rho_{\theta} \; .
\ee

The cosmological solutions for the spatially averaged vacuum
expectation value of $\theta (t)$ depend upon the
initial conditions. If we assume that $\dot{\theta}$ is
initially zero, with $\theta$ taking an arbitrary initial value,
then the solution to (\ref{eqn:NGB1})-(\ref{eqn:NGB3})
is a static universe. If instead we assume that both
$\dot{\theta}$ and $\theta$ have arbitrary initial values,
we get an expansion dominated (initially) by kination:

\be\label{eqn:NGBsol}
\theta(t) = -{\rm ln}\,a(t) \; ,
\quad H = {1\over 3t}\;,
\quad a(t) = t^{1/3} \;,
\ee
where we have taken $f = \sqrt{3}k$ to simplify notation.
Of course in this solution $\rho_{\theta}$  dilutes like $1/a^6$, as
appropriate for kination, \textit{i.e.}, an equation of state dominated by kinetic energy.
Derivative couplings of $\theta$ to ordinary matter will allow
the $\theta$ vacuum energy to be converted to a thermal radiation bath
via decays. Since ordinary matter and radiation dilute like $1/a^3$ and
$1/a^4$, they will eventually dominate the expansion.

Let us interpret this solution in terms of the original global symmetry.
The equation of motion for $\theta$ is just the statement
that the global current is covariantly conserved; the corresponding
conserved global charge of the vacuum $Q$ is proportional to $a^3\dot{\theta}$, which
from (\ref{eqn:NGBsol}) is indeed seen to be a constant.
Once we include matter couplings the vacuum charge $Q$ is no longer constant,
but the total global charge including matter contributions is.

\section{FRW cosmology driven by a pseudo-Nambu-Goldstone boson}

We can modify the discussion in the previous section by introducing
a nonvanishing potential for the $\theta$ field and a noncanonical
kinetic function:
\be\label{eqn:gPNGBaction}
\int d^4x \, \sqrt{-g}\;\left[ \frac {1}{2}F({\rm cos}\,\theta )\, 
g^{\mu\nu}\partial_{\mu}\theta
\partial_{\nu} \theta - V({\rm cos}\,\theta )\right]
\; .
\ee
$F$ and $V$ explicitly break the global $U(1)$ symmetry
down to a discrete periodic remnant:
\be\label{eqn:pshift}
\theta \to \theta + 2\pi N \; ,
\ee
where $N$ is any integer.
This kind of explicit breaking would arise if terms
proportional to powers of $\Phi + \Phi^*$ were present in the original
action. Alternatively, there could be Yukawa couplings or derivative
couplings of $\Phi$ to fermions:
\be
\lambda_{ij} \Phi \,\bar{\psi}_i \psi_j \;,\quad 
\lambda'_{ij} g^{\mu\nu}\partial_{\mu}  
\Phi \,\bar{\psi}_i \gamma_{\nu} \psi_j 
\ee
In this case chiral symmetry breaking will induce,
at the loop level, an effective action of the form
(\ref{eqn:gPNGBaction}).
The simplest possibility, considered in \cite{Frieman:1991tu,Frieman:1995pm},
gives
\be\label{eqn:PNGBaction}
\int d^4x \, \sqrt{-g}\;\left[ \frac {1}{2}f^2\, g^{\mu\nu}\partial_{\mu}\theta
\partial_{\nu} \theta - M^4(1-{\rm cos}\,\theta )\right]
\; ,
\ee
where $M$ is some chiral symmetry breaking scale much smaller than $f$.

Because of this explicit breaking, $\theta$ is now a pseudo-Nambu-Goldstone
boson, with a mass of order $M^2/f$. It is technically natural for this mass
to be small, since it vanishes in a symmetry limit.

Now consider FRW cosmology driven by such a PNGB.
The PNGB equation of motion is:
\be\label{eqn:FHeom}
\ddot{\theta} + 3H\dot{\theta} +{M^4\over f^2}{\rm sin}\,\theta = 0 \;.
\ee

The cosmological solutions again depend upon the initial conditions.
One possibility, considered in \cite{Frieman:1991tu,Frieman:1995pm},
is that $\dot{\theta}$ is
initially zero, with $\theta$ taking an arbitrary initial value.
Then the equation of motion (\ref{eqn:FHeom}) is heavily
overdamped, and $\theta$ remains approximately constant
until $H(t)$ decreases to the point where $H \sim M^2/f$.
Thus the PNGB behaves approximately like dark energy.
The coincidence problem is not solved unless one finds
a rationale for why $M^2/f$, a ratio of two seemingly
independent scales, is roughly equal to the Hubble rate
today. Even then one also needs to explain why the initial value of
$\dot{\theta}$ is much less than $H$.

Another possibility is that both
$\dot{\theta}$ and $\theta$ have generic initial values
during some epoch where
the PNGB dominates the expansion. 
Then for $M/f \ll 1$ the cosmological solution is an
oscillatory perturbation
of the NGB solution (\ref{eqn:NGBsol}). The global charge
of the vacuum $Q \propto a^3\dot{\theta}$ is no longer conserved;
in fact it increases with time like
\be
\frac{\dot{Q}}{Q} \sim t \; .
\ee
This case is more generic than the first one, but does
not provide an explanation of dark energy.

\section{From generic PNGBs to Slinky}

The most general
effective action for the PNGB invariant under the shift
(\ref{eqn:pshift}) is
\be
\label{eqn:genPNGBaction}
\int d^4x \; \sqrt{-g}\; \left[  
\frac{1}{2}\,F({\rm cos}\,\theta )\,P(X)
-V({\rm cos}\,\theta ) 
+{\cal L}_m
\right] \; ,
\ee
where $X=g^{\mu\nu}\partial_{\mu}\theta\partial_{\nu}\theta$,
$F$ and $V$ are arbitrary functions, and $P$ is an arbitrary polynomial.
The matter lagrangian ${\cal L}$ could contain both derivative
couplings to $\theta$ and couplings to functions of cos$\,\theta$.
If there are further explicit breakings of the global symmetry,
the action may also contain a nonperiodic dependence on
$\theta$.

The Slinky model of quintessential inflation introduced in
\cite{Barenboim:2005np}-\cite{Barenboim:2006jh} is a PNGB
quintessence model of the general class just described,
but with some special features.
It does not require tuned initial conditions; in
particular the initial value of $\dot{\theta}$ is of
order $H$.
It is the simplest model with a periodic
equation of state parameter for quintessence.
The quintessence energy density dominates
the Friedmann equation during an earlier epoch
as well as during the present epoch, causing
primordial inflation as well as present-day acceleration.
In order for this to happen,
both the potential and the kinetic energy must
be proportional to the square of the Hubble parameter
times periodic functions:
\be
V \propto H^2 \; , \quad \dot{\theta}^2 \propto H^2 \; .
\ee
The first relation implies that these models will have a PNGB
mass that decreases with time. 

The explicit form of the Slinky is easily derived. 
We begin with the general action (\ref{eqn:genPNGBaction})
and make the simplest nontrivial choice $P(X) = X$.
Without loss of generality, we will take $\theta(t) = 1$ today.
For quintessence the equation of state parameter $w(t)$ should
be close to $-1$ today; for simplicity we will take it to be exactly
$-1$.  Since
\be
1 + w(t) = F({\rm cos}\,\theta)\,\dot{\theta}^2 \; .
\ee
This means that $F({\rm cos}\,\theta)\,\dot{\theta}^2 \to 0$
now. The simplest nonvanishing function that does this is
\be\label{eqn:ourf}
F = {3k^2\over b^2} (1 - {\rm cos}\,\theta)
\; ,
\ee
where $b$ is a dimensionless constant given by
\be
b = \sqrt{\frac{3}{4\pi}}\frac{M_{\rm Planck}}{f} \; .
\ee

The explicit breaking represented by the cosine term in
(\ref{eqn:ourf}) does not have any tunable small parameter
associated with it. Thus one might worry that this breaking is
not small. However since this breaking occurs in the kinetic
function, it generates global charge nonconserving processes
that are suppressed by powers of momenta divided by $k$.
Thus for dynamics well below the Planck scale this breaking is
indeed small.

The PNGB potential $V(\theta)$ must vanish when
we turn off the explicit breaking. Of course this requirement is
a tuning, reflecting the fact that quintessence models do not
explain why the $\theta$-independent cosmological constant
vanishes. In this limit we are kination dominated, with
$w(t) = -{\rm cos}\,\theta = 1$. Thus the simplest  
form for the potential is
\be\label{eqn:ourv}
V \propto (1+{\rm cos}\,\theta)H^2(\theta) \; .
\ee

We need to solve three equations of motion, beginning with the
Friedmann equation:
\be
H^2 = \frac{2}{3k^2}\rho_{\theta} \;.
\ee
The second equation is the scalar EOM:
\be
0 = \ddot{\theta} + 3H\dot{\theta} + \frac{1}{2}\frac{F'}{F}\dot{\theta}^2
+\frac{V'}{F} \; ,
\ee
where a prime denotes a derivative with respect to $\theta$.
The third equation is the continuity equation:
\be
\dot{\rho}_{\theta} = -3(1+w)H\rho_{\theta} \; ,
\ee
where $w(\theta )$ is the equation of state parameter for quintessence.

Plugging in (\ref{eqn:ourf}) and (\ref{eqn:ourv}) we get a solution
with
\be\label{eqn:HPNGB}
H = H_0\;{\rm e}^{\frac{3}{2b}(\theta - {\rm sin}\,\theta )}
\ee
where $H_0$ is the Hubble parameter today. 

Now we have
enough information to specify $F$ and $V$ completely:
\be
F(\theta) &=& {6k^2\over b^2} (1 - {\rm cos}\,\theta) \; ,\nonumber\\
V(\theta) &=& \frac{1}{2}\rho_0 (1+ {\rm cos}\,\theta)
\,{\rm exp}\left[ {3\over b}(\theta - {\rm sin}\,\theta )\right] \; ,
\ee
where $\rho_0 = 3k^2H_0^2/2$ is the analog of $M^4$ in the
standard PNGB quintessence model discussed earlier.
Note that the scalar potential has an additional explicit
breaking that does not respect the periodicity (\ref{eqn:pshift}).
This breaking is small during any epoch such that
$H/k \ll 1$.
 
Slinky is a PNGB quintessence model characterized by a periodic
equation of state $ w = -{\rm cos}\,\theta $.
The effective mass of the PNGB is tied to the expansion rate,
and therefore decreases over time. These two features combined
allow the same PNGB to serve the role both of the inflaton in 
the early universe and the quintessence field that drives the accelerated
expansion in the current epoch. 

The number of inflationary epochs that have occurred is controlled
by the dimensionless parameter $b$; larger values of $b$ mean
more inflationary periods. Most values of $b$ are ruled out by
the requirement that the universe had at most a tiny inflationary
component during BBN \cite{Bean:2001wt}.
Note from (\ref{eqn:ourf}) 
that $b \lsim 0.49$ implies that $f > M_{\rm Planck}$. The
Slinky models considered in \cite{Barenboim:2006rx} have values
in the range $0.09 < b < 0.4$. This does not necessarily mean
that these models require trans-Planckian vevs; for example,
if we replace the single PNGB by $N$ identical ones, then
$b$ is effectively rescaled as $b \to b/\sqrt{N}$. Thus small
values for $b$ do not indicate a scenario that is technically
out of control.

\begin{figure}
\centerline{\epsfxsize 5.4 truein \epsfbox {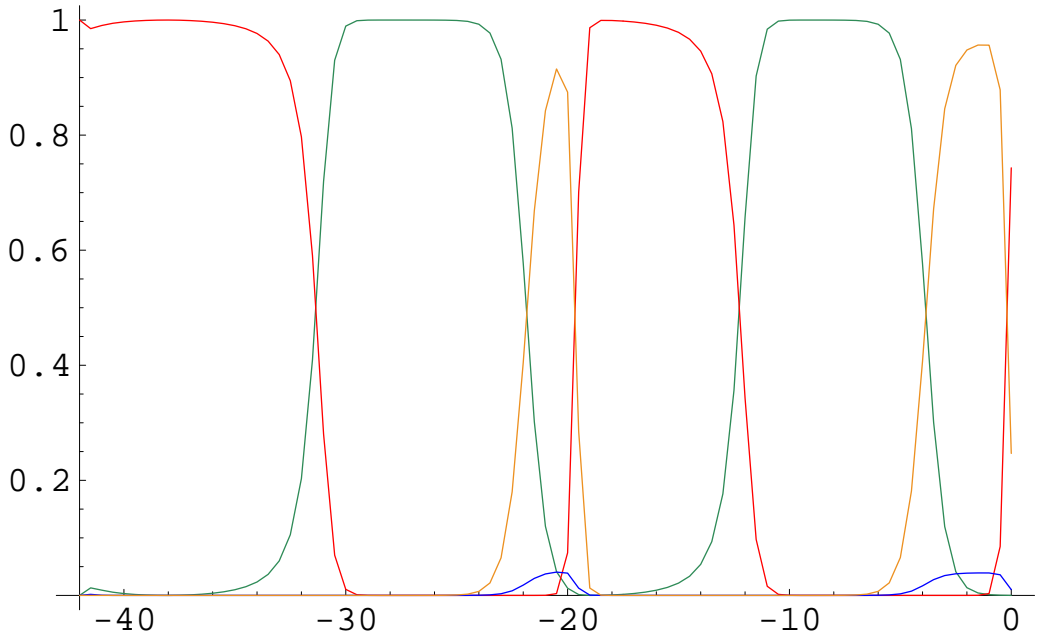}}
\caption{\label{fig:omegas} The cosmological history of
the simplest Slinky model. The x-axis is the logarithm base 10
of the scale factor. Shown are energy densities of
radiation (green), dark matter (orange), baryons (blue)
and dark energy (red), as a fraction of the critical density.
\hbox to140pt{}}
\end{figure}

The nonstandard cosmological history of this model is
depicted in Figure \ref{fig:omegas}. 
The energy densities of radiation,
dark matter, baryons and dark energy, computed as a fraction
of the critical density, are plotted as function of the
logarithm of the scale factor. 
We have chosen to start the cosmological evolution at a
scale factor of $10^{-42}$, but this choice is not essential.
We have assumed that the Friedmann equation is initially
dominated by the PNGB. For simplicity we set the initial
radiation density to zero, so all radiation arises from
PNGB decays, as explained in the next section. 

\begin{figure}
\centerline{\epsfxsize 5.4 truein \epsfbox {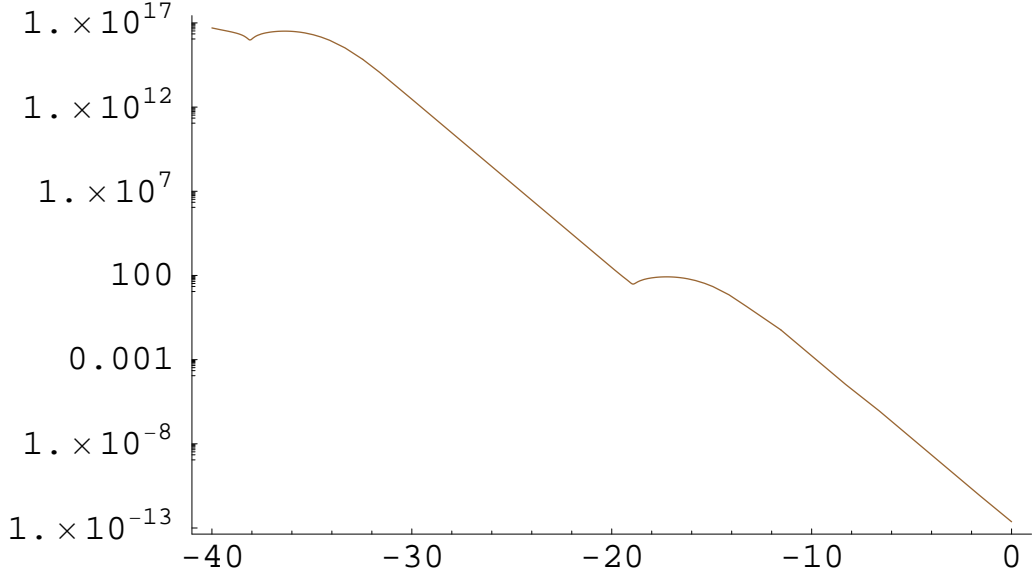}}
\caption{\label{fig:p1} The cosmological history of
the simplest Slinky model. The x-axis is the logarithm base 10
of the scale factor. Shown are energy densities of
radiation (green), dark matter (orange), baryons (blue)
and dark energy (red), as a fraction of the critical density.
\hbox to140pt{}}
\end{figure}

The radiation temperature as
a function of the scale factor is shown in Figure \ref{fig:p1}.
During the initial inflation the temperature is
roughly constant at around $10^{16}$ GeV, increasing mildly
towards the end of this first inflationary epoch. This
behavior is in contrast to standard primordial inflation,
where the temperature first decreases very rapidly, then
increases very rapidly (reheating). The new behavior
is due to the time-varying equation of state.

\section{Couplings to matter}

The allowed couplings of $\theta$ to matter can be divided
into two classes:
\begin{enumerate}
\item Couplings of functions of cos$\,\theta$ to matter.
These couplings explicitly break the global $U(1)$ symmetry.
It is technically natural to take the dimensionless
coupling constants for all such couplings to be small.
\item Derivative couplings of $\theta$ to matter.
Some of these couplings respect the full global $U(1)$ symmetry,
so there is no symmetry argument for tuning their
dimensionless coupling constants to be small; indeed
we will assume that they are of order one. However almost
all such couplings are higher dimension operators suppressed
by powers of momenta over powers of $M_{\rm Planck}$. 
\end{enumerate}

Thus for Slinky, as well as for more generic
PNGB quintessence models, it is a good approximation to only
include couplings to matter of the form
\be\label{eqn:mattercoup}
\lambda_{ij} \;g^{\mu\nu}\partial_{\mu}  
\theta \,\bar{\psi}_i \gamma_{\nu} \psi_j \; ,
\ee
where $\lambda_{ij}$ is dimensionless.
A caveat is that the Slinky potential (\ref{eqn:ourv})
has an additional explicit breaking of the
global $U(1)$ symmetry, that we are assuming has no
analog in the matter couplings.

Couplings of the form (\ref{eqn:mattercoup}) allow the
$\theta$ field to decay into ordinary matter.
This process can be modeled, 
albeit roughly \cite{Dolgov:1996qq}, as
an additional friction term in the $\theta$ equation
of motion:
\be
0 = \ddot{\theta} + 3H\dot{\theta} + \Gamma\dot{\theta}
+ \frac{1}{2}\frac{F'}{F}\dot{\theta}^2
+\frac{V'}{F} \; .
\ee
Here $\Gamma$ is the decay width, which for
decays into pairs of fermions can be generically
written
\be
\Gamma = k_m\,m_{\rm eff}(t)  \; ,
\ee
where $k_m \sim \lambda_{ij}$ are dimensionless couplings
which we will take to be approximately constant 
with magnitudes in the range
$.1$ to $.01$;
$m_{\rm eff} \propto H$ is the time varying
mass of the PNGB, obtained by expanding the
potential (\ref{eqn:ourv}).

This process converts quintessence vacuum energy
into matter and radiation. Thus Slinky produces the
following nonstandard cosmological history:
\begin{itemize}
\item During a primordial epoch, $w$ is close to
-1 and inflation occurs. 
\item Eventually $w$ increases
towards +1, and the radiation produced from $\theta$
decays, which is now diluting less rapidly than
the quintessence energy, comes to dominate the expansion.
\item The process repeats. In the original
Slinky model \cite{Barenboim:2005np}, which
we will use from now on, the second radiation
dominated epoch overlaps with the time of Big Bang
Nucleosynthesis. 
\item Well before BBN time, $m_{\rm eff}$
becomes so small that the PNGB decays to matter effectively
turn themselves off, due to kinematic suppression.
PNGB decays to photons continue, but these are loop-suppressed
and so do not give large entropy production.
\item Right now we are entering the third inflationary
phase.
\end{itemize}

From the arguments given above, the only couplings in this
model whose values affect the cosmological evolutions are
$b$ and the $\lambda_{ij}$. The value of $b$ is coarsely
adjusted such that
we ensure that BBN time is radiation dominated and that
$w$ is close to $-1$ today. The dominant $\lambda_{ij}$
couplings are adjusted such that the radiation and matter
fractions
$\Omega_r/\Omega_{\Lambda}$ and  $\Omega_{DM}/\Omega_{\Lambda}$
come out to their measured values today.
For simplicity we assume that all of the dark matter is
produced thermally, as \textit{e.g.} in standard WIMP
scenarios.
The remaining matter couplings (for the moment) remain free.

We now see more clearly why the
the temperature history represented in Figures \ref{fig:omegas}
and \ref{fig:p1} is so nonstandard.
The coupling between matter and the PNGB field
force them track each other, giving significant entropy production
even after the first period of inflation . This entropy production
turns off well before BBN time due to the kinematic suppression of
PNGB decays. 

\section{Baryogenesis}

If baryon number $B$ and lepton number $L$ are exactly
conserved, then $B$ or $L$ violating decays of the
PNGB simply do not occur. This can be seen
from the coupling (\ref{eqn:mattercoup}), which with
integration by parts vanishes for any conserved current
$j_{\mu} = \bar{\psi}\gamma_{\mu}\psi$.

However we certainly do not expect the global symmetries
$B$ and $L$ to be respected by unification scale physics.
We therefore introduce all possible $B$ and $L$ violating
dimension six operators:
\be
\frac{\psi_i\psi_j\psi_k\psi_l}{\Lambda_{ijkl}^2}
\ee
suppressed by some superheavy scales $\Lambda_{ijkl}$.
Suppressing flavor labels, this implies that the
$B$ or $L$ violating decay width of the PNGB can be written
\be
\Gamma = \lambda^2 \; {m_{\rm eff}^5 \over \Lambda^4}
\; .
\ee

Thus PNGB evolution and decay during the first inflationary epoch
will produce a net $B$ and $L$ asymmetry, including a
$B-L$ asymmetry that will survive the $B+L$ washout by
sphalerons during the electroweak phase transition.

\begin{figure}
\centerline{\epsfxsize 5.4 truein \epsfbox {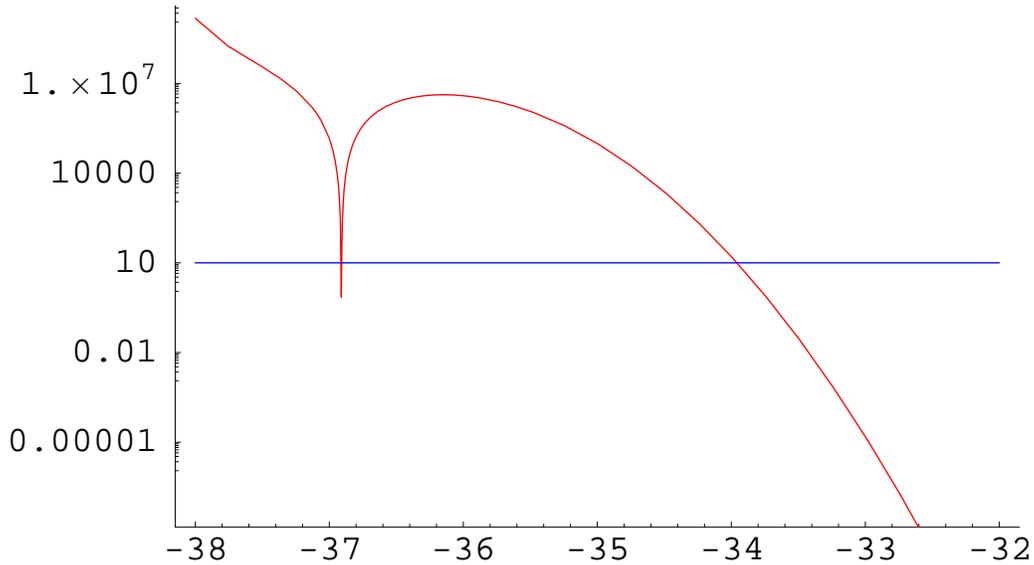}}
\caption{\label{fig:equil} The ratio of the relative rates
of $B$ or $L$ violating processes versus the cosmological
evolution of the PNGB field, as a function of the logarithm base 10
of the scale factor. We have taken $\Lambda = 10^{15}$ GeV.
The ratio is shown in red;
$B$ or $L$ violation goes out of equilibrium below
the blue line.
\hbox to140pt{}}
\end{figure}

At sufficiently high temperatures the $B$ or $L$ violating
processes will be in thermal equilibrium. The equilibrium
condition is \cite{Cohen:1987vi}
\be
\Gamma \gg {\ddot\theta\over \dot\theta} = \frac{3}{2}
(1-{\rm cos}\,\theta)H_{PNGB} \; ,
\ee
where $H_{PNGB}$ denotes the Hubble rate that would
result from the PNGB alone, given in equation (\ref{eqn:HPNGB}).
Figure (\ref{fig:equil}) shows an estimate of the value of the
scale factor at which $B$ or $L$ violating processes decouple,
assuming $\Lambda = 10^{15}$ GeV. The corresponding
decoupling temperature is about $8\times 10^{15}$ GeV.

Once we are out of equilibrium the baryon asymmetry
continues to grow via PNGB decays. The net
baryon number density as a function of time can
be written as \cite{Cohen:1987vi}
\be\label{eqn:nbdot}
\dot{n}_B = \lambda  \; {m_{\rm eff}^5 \over \Lambda^4} 
\; { 12 k^2\over b^2} \;
\dot{\theta}\;{\rm sin}\,\theta
\; .
\ee
It is useful to 
express the baryon number density as a function of $\theta$, \textit{i.e.}
\be
n^\prime_B =  \lambda  \; {m_{\rm eff}^5 \over \Lambda^4} \; 
{12 k^2\over b^2}  \;
{\rm sin}\,\theta  
\;. 
\ee

To calculate the baryon asymmetry produced, we have to include
the expansion of the universe. This can be
easily done by replacing all the volume factors in the equations above
by the corresponding comoving volumes, proportional to $a^3$,
\be
\frac{d}{d\theta} \; ( a^3 n_B ) =  \lambda  
\; a^3 \;  {m_{\rm eff}^5 \over \Lambda^4} \; { 12 k^2\over b^2} \;
 {\rm sin}\,\theta  
\; .
\ee 
To obtain the net baryon number density we have to numerically integrate this 
expression from the decoupling temperature 
to $\theta_f$, the temperature where 
baryon production from PNGB decays is cut off kinematically
by the decrease in $m_{eff}$.
For decays to baryons $\theta_f$ corresponds to a scale factor
of approximately $10^{-27}$, when $m_{\rm eff} \sim H(\theta) \simeq 1$ GeV. 

From the baryon density we want to extract the baryon 
to photon ratio, $\eta$. As usual the photon number density
is given by 
\be
n_\gamma = {2 \zeta(3) \over \pi^2} \; T^3 \; . 
\ee

Contrary to the standard scenarios \cite{Buchmuller:2005eh,Trodden:2004mj},
where the baryon (more precisely, $B-L$)
to photon ratio does not change after baryogenesis, in our model  
there is significant 
entropy production
due to the coupling of the PNGB field to radiation. The net effect of such
a production will be to dilute any baryon to photon ratio produced 
before BBN, where entropy production stops. Therefore $\eta$ can 
still be calculated at $\theta_f $, but we have to include the extra 
dilution factor $\gamma$, given by
\be
\gamma ={ S_{\theta_f} \over S_{\theta_{BBN}}} \; .
\ee
Here $S = g_* \; a^3 \;  T^3$, with $g_*$ the effective number of relativistic
degrees of freedom.
Alternatively, we can calculate $\eta$ directly at BBN time.

The only input parameters left to adjust in our simple model
are the scale $\Lambda$ and the dimensionless coupling $\lambda$. 
We consider two well-motivated scenarios.
In the first scenario,
$\Lambda \simeq 10^{15}$ GeV, just about saturating the generic
lower bounds from the nonobservation of proton decay \cite{Nath:2006ut}.
We estimate the baryon to photon ratio $\eta$ in this case
to be
\be 
\eta \simeq  \; 4 \lambda \;  \times 10^{-10} \; .
\ee
This gives the observed baryon asymmetry for $\lambda \simeq 1$.

The second scenario has $\Lambda \simeq 10^{11}$ GeV, as might
be appropriate in a leptogenesis model. Here we estimate
\be 
\eta \simeq  \; 5 \lambda \;  \times 10^{-8} \; ,
\ee
which gives the observed baryon asymmetry for 
$\lambda \simeq 0.01$.

The cosmological history of $\eta$ is unlike that of
previously discussed models of quintessential
baryogenesis \cite{DeFelice:2002ir}-\cite{Carroll:2005dj}.
At early times the PNGB effective mass is large, and the baryon number violating
processes are least suppressed. 
A baryon or lepton asymmetry is produced in thermal
equilibrium, but this contribution is negligible compared
to the asymmetry produced later by PNGB decays. This
is basically because the PNGB decay process is enhanced by
$f^2$, the square of the Planckian scalar vev, as seen in
(\ref{eqn:nbdot}).
The net baryon production turns itself off kinematically
as $m_{eff}$ decreases. The baryon to photon ratio
drops dramatically
during the subsequent second inflation era, where there is further
large entropy production.   
The fact that the baryon to photon number density observed
today is very small is thus due entirely to the existence
of a second inflation era; $B$ or $L$ violation was only mildly
suppressed at the time that most of the net baryon or lepton
excess was created.

\section{Conclusions}
We have shown that a single noncanonical PNGB could be
responsible for both primordial inflation and the present day accelerated
expansion, while simultaneously generating the observed baryon excess.
The baryon asymmetry
is generated after the first inflationary period, from PNGB decays via
dimension six operators that violate $B$ and/or $L$.
The ratio of baryon to photon number densities is
greatly diluted later on, via entropy production from
$B$ and $L$ conserving PNGB decays. 
The baryon energy density is completely
negligible expect during two eras: the present day and an era
around the time of the electroweak phase transition.

While economical, technically natural, and consistent with current data,
our model has some theoretical shortcomings. It explains
dark energy but does not solve the cosmological constant problem.
There is also no first principles explanation for why the
PNGB potential scales like the square of the Hubble rate.
Resolving these shortcomings presumably involves an
ultraviolet completion of the PNGB effective theory into
a more fundamental framework.

One of the distinguishing features of this model, 
is that it predicts substantial entropy production
in the era between the electroweak phase transition and BBN.
If dark matter is predominately composed of thermally produced
WIMPs, this prediction can be tested by combining collider
data with signals from direct and indirect 
WIMP searches \cite{Barenboim:2006jh},
\cite{Kamionkowski:1990ni}-\cite{Gelmini:2006pq}.

Because of the entropy production after the second
inflation era (but before BBN), any pre-existing  baryon
asymmetry is much diluted. This requires that the
operators responsible for $B$ violation be not too
much suppressed, forcing us close to the current
experimental bounds for proton decay. Thus another
prediction is that proton decay will be observed in
one of the future proposed 
experiments \cite{Nakamura:2003hk}-\cite{de Bellefon:2006vq}.

Future dark energy probes will pin down the
equation of state of dark energy with much greater
precision. In our simple model we artificially set
$w=1$ exactly at redshift $z=0$; more relevant is
that $w$ varies by about 2\%\ as redshift is varied
between 0 and 2. This variation is comparable to the
one-sigma projected combined errors after the Stage IV dark
energy probes \cite{Albrecht:2006um}.

Models of the type discussed here predict running
of the spectral indices of the Cosmic Microwave Background
(CMB) \cite{Barenboim:2007ii}. These effects may be
large enough to extract from observations of the PLANCK
satellite \cite{Pahud:2007gi}. The smoking gun of a
predictive model of quintessential inflation is that
these CMB effects, a result of inflation, are
directly related to the detailed equation of state
of dark energy.

\subsection*{Acknowledgments}
The authors are grateful to Bill Bardeen and Scott Dodelson
for useful discussions. 
GB acknowledges support from the Spanish MEC and FEDER under 
Contract FPA 2005/1678,
and the Generalitat Valenciana under Contract GV05/267.
JL acknowledges the Aspen Center for Physics, where part of this
work was completed.
Fermilab is operated by the Fermi Research Alliance LLC under
contact DE-AC02-07CH11359 with the U.S. Dept. of Energy.

\newpage


\end{document}